\documentclass[structabstract]{aa}  

\usepackage{graphicx}
\usepackage{txfonts}
\usepackage{lscape}
\usepackage{natbib}
\usepackage{rotating}

\def\ie{i.e.}

\def\deg{\ifmmode^\circ\else$^\circ$\fi}
\def\alphaTF{\ifmmode{\alpha_{\mathrm{\,{\small TF}}}}\else{$\alpha_{\mathrm{\,{\small TF}}}$}\fi}

\begin{document}

\title{Evolution along the sequence of S0 Hubble types\\induced by dry minor mergers}
\subtitle{I - Global bulge-to-disk structural relations}

\author{M.~Carmen Eliche-Moral\inst{1}, A.~C\'esar Gonz\'{a}lez-Garc\'{\i}a\inst{2,3}, J.~Alfonso L.~Aguerri\inst{2,3}, Jes\'us Gallego\inst{1}, Jaime Zamorano\inst{1}, Marc Balcells\inst{4,2,3}, \& Mercedes Prieto\inst{2,3}}

\institute{Departamento de Astrof\'{\i}sica, Universidad Complutense de Madrid, E-28040 Madrid, Spain,  \email{mceliche@fis.ucm.es}
  \and
Instituto de Astrof\'{\i}sica de Canarias, C/ V\'{\i}a L\'actea, E-38200 La Laguna, Tenerife, Spain
  \and
Departamento de Astrof\'{\i}sica, Universidad de La Laguna, E-38200 La Laguna, Tenerife, Spain
  \and
Isaac Newton Group of Telescopes, Apartado 321, E-38700 Santa Cruz de La Palma, Canary Islands, Spain
}

   \date{Received December 22, 2011; accepted September 4, 2012}

\abstract{
Recent studies have argued that galaxy mergers are not important drivers for the evolution of S0's, on the basis that mergers cannot preserve the coupling between the bulge and disk scale-lengths observed in these galaxies and the lack of correlation of their ratio with the S0 Hubble type. However, about 70\% of present-day S0's reside in groups, an environment where mergers and tidal interactions dominate galaxy evolution.
}{
We investigate whether the remnants resulting from collision-less N-body simulations of intermediate and minor mergers onto S0 galaxies evolve fulfilling global structural relations observed between the bulges and disks of these galaxies.
}{
Different initial bulge-to-disk ratios of the primary S0 have been considered, as well as different satellite densities, mass ratios, and orbits of the encounter. We have analysed the final morphology of the remnants in images simulating the typical observing conditions of S0 surveys. We derive bulge$+$disk decompositions of the final remnants to compare their global bulge-to-disk structure with observations. 
}{
We show that all remnants present undisturbed S0 morphologies according to the prescriptions of specialized surveys. The dry intermediate and minor mergers induce noticeable bulge growth (S0c$\longrightarrow$S0b and S0b$\longrightarrow$S0a), but affect negligibly to the bulge and disk scale-lengths. Therefore, if a coupling between these two components exists prior to the merger, the encounter does not break this coupling. This fact provides a simple explanation for the lack of correlation between the ratio of bulge and disk scale-lengths and the S0 Hubble type reported by observations.  
}{
These models prove that dry intermediate and minor mergers can induce global structural evolution within the sequence of S0 Hubble types compatible with observations, meaning that these processes should not be discarded from the evolutionary scenarios of S0's just on the basis of the strong coupling observed between the bulge and disk scale-lengths in these galaxies.
}


\titlerunning{Evolution along the sequence of S0 Hubble types through dry minor mergers. I}
\authorrunning{Eliche-Moral et al.}

   \maketitle

\section{Introduction}
\label{Sec:introduction}

Lenticular galaxies (S0's) are disk galaxies with a smooth appearance due to their lack of spiral arms and star-forming regions \citep{1926ApJ....64..321H,1959HDP....53..275D,1961hag..book.....S}. The traditional picture of these galaxies as a transition class between spirals and ellipticals is drastically changing in the last years, because although S0's tend to exhibit massive central bulges, recent observations show that their bulge-to-disk ratios ($B/D$) can span a wide range of values. In this sense,  \citet[][]{2007ApJ...665.1104B} found no correlation between $B/D$ and the typical sequence of Hubble types (S0-Sa-Sb-Sc): their S0 galaxies had a mean $B/D$ value similar to that of late-type spirals ($B/D \sim 0.25$). Studies using larger samples of galaxies including lenticulars show that, although the bulge-to-total luminosity ratio ($B/T$) increases from late-type to early-type galaxies, some S0 galaxies can exhibit $B/T$ values similar to later types \citep[$B/T \approx 0.1-0.3$, see][L10 henceforth]{2009ApJ...696..411W,2010MNRAS.405.1089L}. This fact has made three different groups independently to update the van den Bergh's version of the Hubble tuning fork recently, locating S0 galaxies in a sequence (S0a-S0b-S0c) parallel to the one of spirals (Sa-Sb-Sc), instead of placing them at an intermediate location between ellipticals and spirals, as done in the traditional Hubble diagram \citep[see L10;][]{2011MNRAS.418.1452L,2011MNRAS.416.1680C,2012ApJS..198....2K}. 

The large dispersion of $B/T$ for S0 galaxies may be reflecting different formation and evolution mechanisms, as suggested by \citet[][]{1976ApJ...206..883V}. The later appearance of S0's into the cosmic scenario and the noticeable rise of their population in clusters at the expense of spirals since $z\sim 0.5$ evidence that S0's must derive from them \citep[][L10]{1996A&A...313...45D,1996ApJ...457L..73C,2000ApJ...542..673F,2001AJ....121..820G,2005ApJ...620...78S,2005ApJ...623..721P,2007ApJ...665.1104B,2008ApJ...684..888P,2009ApJ...697L.137P,2009ApJ...692..298W}. However, the main mechanisms after their formation and evolution are still unsettled, with data indicating that quite different processes may have contributed to it depending on the environment and on their luminosity \citep{2007ApJ...661L..37B,2009MNRAS.394.1991B,2009AJ....138..579K,2009A&A...508.1141S,2010A&A...515A...3H}. This has been pointed out in a recent review about S0 galaxies by \citet[][]{2012AdAst2012E..28A}, who derives this conclusion attending to the different properties of the photometric components of lenticular galaxies and their implications in the formation and evolution mechanisms of these galaxies. 

S0's, like most disk galaxies, show a strong coupling between the bulge effective radius  ($r_\mathrm{e}$) and the disk scale length \citep[$h_\mathrm{D}$, see][L10]{2004A&A...414..905H,2009ApJ...692L..34L}. As argued by \citeauthor{2009ApJ...692L..34L}, the hierarchical merger scenario cannot account for such coupling, because bulges grow by mergers whereas disks grow later from the accretion of left-over gas, these two processes being thus decoupled \citep[L10]{2009ApJ...692L..34L}.  This reasoning has left mergers in the backstage of the S0's evolution scenarios, favouring other processes instead, such as ram pressure stripping, internal secular evolution, or simple fading of the stellar populations \citep[see][]{1980ApJ...236..351D,1999MNRAS.308..947A,1999MNRAS.304..465M,2000Sci...288.1617Q,2005MNRAS.363..496A,2005ApJ...620...78S,2006ARep...50..785T,2009MNRAS.400.1706A,2009MNRAS.394...67A}. 

Concerning simple fading, some studies support that this process alone cannot explain the conversion of spirals into S0's in clusters, as it requires the enhancement of the bulge luminosity compared to the one of the disk \citep[not just its fading, see][]{2004ApJ...616..192C}. The relevance of gas stripping in clusters is observationally proven \citep{1999AJ....117..181K,2007ApJ...659L.115C}, but this process is not sufficient to explain the origin of S0's either. First, most S0's ($\sim 70$\%) reside in groups, not in clusters, thus ruling out this formation mechanism for the majority of them (see references below). Moreover, \citet{2009ApJ...692L..34L} argue that the stripping hypothesis (including quiescent star formation) cannot explain why S0's have a significantly smaller fraction of bars (46\%) than S0/a galaxies (93\%) or spirals (64\%-69\%), which are their assumed progenitors \citep[see also][]{2009A&A...495..491A,2011MNRAS.410L..18B}. 

Bars themselves and internal secular processes are good candidates for explaining the formation and evolution of S0's, as they guarantee both the enhancement of the bulge luminosity (inducing star formation in the centre) and a tight structural binding between the bulge and the disk (thanks to the inflow of disk material towards the galaxy centre through the bar). This seems to be supported by the finding of that $\sim 97$\% of S0's have ovals, which might be bar relics \citep{2009ApJ...692L..34L}. However, although models of internal secular evolution demonstrate that this mechanism can give rise to bulges of structural properties compatible with observations of spirals, gas-free models (which would be the candidates to explain the formation of the S0 bulges) produce bulges with too low concentrations and too large compared to real ones. Observations indicate that S0's have a S\'ersic index $n\sim 2$ and $r_\mathrm{e}/h_\mathrm{D}\sim 0.2$ on average (L10), whereas these models buildup bulges with $n\lesssim 1$ and $r_\mathrm{e}/h_\mathrm{D}\sim 0.4$ \citep[][]{2006ApJ...645..209D}.

The environment of S0's provide clues that mergers must influence their evolution. At $z<0.5$, the fraction of S0's in groups is similar to that of clusters \citep{2009ApJ...692..298W}. Considering that most galaxies reside in groups \citep[$\sim 70$\%, see][]{1982ApJ...257..423H,2006ApJS..167....1B,2007ApJ...655..790C}, this means that the majority of S0's in the Universe are located in groups (not in clusters). Moreover, evolved groups have a much higher fraction of S0's than young groups \citep[][]{2009AJ....138..295F,2010ApJ...713..637B}. This morphological transformation from spirals towards S0's in groups is obviously driven by the processes governing galaxy evolution in these environments, which are mergers and tidal interactions. Additionally, at least $\sim 16$\% of normal relaxed S0's exhibit tidal debris, again suggesting past merging \citep{2012ApJ...753...43K}, and recent studies show that the low-level star formation activity in early-type galaxies (E-S0's) at intermediate redshifts is mostly driven by minor mergers too \citep{2009MNRAS.394.1713K,2011MNRAS.411.2148K}. The spread in the ages of the globular clusters in S0's compared to ellipticals might be also pointing to a relevant impact of minor mergers in their formation, as these processes could bring bluer (and on average younger) globular clusters with them from the smaller hosts that could explain it \citep{2011A&A...525A..20C}. It is thus straightforward to start questioning the traditional picture of mergers as catastrophic events, incapable of reproducing the structural bulge-disk binding observed in S0's. 

Simulations show that major mergers (i.e., encounters with mass ratios above 1:3) build up ellipticals, not S0-like systems, unless the event involves an extraordinary gas amount or a large bulge in the progenitors \citep{2005MNRAS.357..753G,2006MNRAS.369..625N,2007ApJ...658...60B}. But even in this case, the $r_\mathrm{e}/h_\mathrm{D}$ ratios are too high ($\sim 0.4$) compared to the typical ones of real S0's ($\sim 0.2$, see L10) and the remnants lie well below observational data in the $r_\mathrm{e}$-$h_\mathrm{D}$ plane. However, intermediate mergers (i.e., events with mass ratios between 1:4 and 1:7) can produce remnants that photometrically and kinematically resemble S0 galaxies \citep[][]{1998ApJ...502L.133B,2005A&A...437...69B}. 

Although simulations suggest that minor-to-intermediate mergers are capable of reproducing the observed distribution of galaxies in the S\'ersic index ($n$) vs.\, $B/D$ plane, very few studies have compared the distributions of the bulge and disk scale-lengths with observations \citep[][]{2001A&A...367..428A,2003MNRAS.338..880S,2006A&A...457...91E}. In particular, \citet{2003MNRAS.338..880S} studied the effects of mergers on the structural properties of disk-like systems in a cosmological context, and found that the progenitors of these objects at different redshifts exhibited global structures compatible with real galaxies, whereas their counterparts at $z=0$ do not (compare their Figs.\,2 and 5). If intermediate and minor mergers have played any role in the evolution of S0's, the remnants of merger simulations should not just be compatible with the distribution of real S0's in the $r_\mathrm{e}$-$h_\mathrm{D}$ plane, but they should also reproduce the lack of correlation found in S0's between $r_\mathrm{e}/h_\mathrm{D}$ and $n$. The analogous result for spirals has been interpreted as an evidence of the fact that the Hubble Sequence is scale free \citep[][]{1996ApJ...457L..73C,1996A&A...313...45D,2001AJ....121..820G,2004A&A...414..905H,2007ApJ...665.1104B}. 

Therefore, we have re-addressed the question of whether intermediate and minor mergers can induce evolution compatible with the bulge-to-disk coupling observed in real S0's or not. The aim of the present paper is to demonstrate that there is no reason to exclude dry minor mergers as a valid mechanism for driving the evolution of lenticular galaxies along the S0 sequence. We have analysed the evolution induced by gas-free intermediate and minor merger events onto S0 galaxies, with mass ratios ranging from 1:6 to 1:18. 

The paper is structured as follows. The models are briefly described in Sect.~\ref{Sec:models}. The morphology of the final remnants is analysed through the simulation of realistic images in Sect.~\ref{Sec:imageSimulations}. In Sect.~\ref{Sec:photometricParameters}, we describe the bulge-disk decompositions performed to them. The structural analysis of the remnants and the comparison with real S0's are presented in Sect.~\ref{Sec:results}. The discussion of the results can be found in Sect.~\ref{Sec:discussion}. Final conclusions are addressed in Sect.~\ref{Sec:conclusions}. A concordant cosmology is assumed for image simulations of galaxies at different cosmological distances \citep[$\Omega_\mathrm{M} =0.3$, $\Omega_\Lambda =0.7$, $H_0 = 70$\,km\,s$^{-1}$\,Mpc$^{-1}$, see][]{2007ApJS..170..377S}. All magnitudes used in the paper are in the Vega system.

\begin{table*}
\begin{minipage}[t]{\textwidth}
\caption{Orbital and scaling parameters of each merger experiment}
\label{tab:models}
\centering
\begin{tabular}{llccrcc}
\hline\hline
\multicolumn{1}{c}{Model code} & \multicolumn{1}{c}{EM06 code} &  $M_{\mathrm{2}}/M_{\mathrm{1}}$  & $R_{\mathrm{pericentre}}/h_\mathrm{D}$  & \multicolumn{1}{c}{$\theta _\mathrm{1}$ ($^o$)} &  \multicolumn{1}{c}{\textrm{Primary $B/D$}} & $\alphaTF$ \\
\multicolumn{1}{c}{(1)}    & \multicolumn{1}{c}{(2)} & \multicolumn{1}{c}{(3)}         & \multicolumn{1}{c}{(4)}    & \multicolumn{1}{c}{(5)}        & \multicolumn{1}{c}{(6)}      & \multicolumn{1}{c}{(7)}     \vspace{0.05cm}\\\hline\vspace{-0.3cm}\\
 (a)  M6 Ps Db      & M2TF35D &  1:6 (M6)   &  0.73 (Ps) &  30 (D) & \textrm{0.5} (b)  & 3.5 \\
 (a2) M6 Ps Db TF3  & M2TF3D  &  1:6 (M6)   &  0.73 (Ps) &  30 (D) & \textrm{0.5} (b)  & 3.0 \\
 (a3) M6 Ps Db TF4  & M2TF4D  &  1:6 (M6)   &  0.73 (Ps) &  30 (D) & \textrm{0.5} (b)  & 4.0 \\
 (b) M6 Ps Rb       & M2R     &  1:6 (M6)   &  0.73 (Ps) & 150 (R) & \textrm{0.5} (b)  & 3.5  \\
 (c) M6 Pl Db   & \multicolumn{1}{l}{...} &  1:6 (M6) &  8.25 (Pl) &  30 (D)   & \textrm{0.5} (b)  & 3.5 \\
 (d) M6 Pl Rb   & \multicolumn{1}{l}{...} &  1:6 (M6) &  8.25 (Pl) &  150 (R)  & \textrm{0.5} (b)  & 3.5 \\
 (e) M6 Ps Ds & \multicolumn{1}{l}{...}   &  1:6 (M6) &  0.87 (Ps) &  30 (D)   & \textrm{0.08} (s) & 3.5 \\
 (f) M6 Ps Rs & \multicolumn{1}{l}{...}   &  1:6 (M6) &  0.87 (Ps) & 150 (R)   & \textrm{0.08} (s) & 3.5\\ \vspace{-0.4cm}\\\hline\vspace{-0.3cm}\\
 (g) M9 Ps Db & M3TF35D & 1:9 (M9) &  0.79 (Ps) &  30 (D)   &  \textrm{0.5} (b)  & 3.5 \\
 (g2) M9 Ps Db TF3 & M3TF3D & 1:9 (M9) &  0.79 (Ps) &  30 (D)   &  \textrm{0.5} (b)  & 3.0 \\
 (g3) M9 Ps Db TF4 & M3TF4D & 1:9 (M9) &  0.79 (Ps) &  30 (D)   &  \textrm{0.5} (b)  & 4.0 \\
 (h) M9 Ps Rb      & M3R    & 1:9 (M9) &  0.79 (Ps) & 150 (R)  & \textrm{0.5} (b)  & 3.5\\ \vspace{-0.4cm}\\\hline\vspace{-0.3cm}\\
 (i) M18 Ps Db & M6TF35D & 1:18 (M18)&  0.86 (Ps) &  30 (D)   & \textrm{0.5} (b)  & 3.5 \\
 (j) M18 Ps Rb & M6R     & 1:18 (M18)&  0.86 (Ps) & 150 (R)   & \textrm{0.5} (b)  & 3.5 \\
 (k) M18 Pl Db & \multicolumn{1}{l}{...} &  1:18 (M18)& 8.19 (Pl) &  30 (D)  & \textrm{0.5} (b)  & 3.5\\
 (l) M18 Pl Rb & \multicolumn{1}{l}{...} &  1:18 (M18)& 8.19 (Pl) &  150 (R) & \textrm{0.5} (b)  & 3.5 \\\hline\\
\end{tabular}
\begin{minipage}[t]{\textwidth}{\small
\emph{Columns}: (1) Model code: M$m$P[l/s][D/R][b/s][TF3/4], consult the text. The identification letters in parentheses for each model are the same as those used in EM11. The models with $\alphaTF=3.0$ or 4.0 from EM06 have been labelled with letters "a2", "a3", "g2", and "g3" to refer them to their analogues with $\alphaTF=3.5$. (2) Model code in EM06, for those models that were already presented in that paper. (3) Luminous mass ratio between satellite and primary galaxy. (4) First pericentre distance of the orbit, in units of the original primary disk scale-length. (5) Initial angle between the orbital momentum and the primary disk spin. This angle determines whether the orbit is prograde (direct) or retrograde. (6) Bulge-to-disk ratio of the original primary galaxy used in the experiment: $B/D=0.5$ (big bulge) or $B/D=0.08$ (small bulge). (7) Value of $\alphaTF$ assumed for the scaling of the satellite to the primary galaxy.}
\end{minipage}
\end{minipage}
\end{table*}

\section{Description of the models} 
\label{Sec:models}

We have analysed the battery of collision-less simulations of intermediate and minor mergers described in \citet{2006ApJ...639..644E} and \citet[]{2011A&A...533A.104E} (EM06 and EM11 respectively, henceforth), in which both primary and satellite are modelled as disk-bulge-halo galaxies with realistic density ratios (in total, 16 models).  Our minor merger experiments differ in the considered orbital pericentre distances, the spin-orbit coupling, the primary-to-satellite luminous mass ratios, the central density ratios, and the galaxy types for the initial primary galaxy. As these models have been extensively described in the original papers, we just provide a brief description here, asking the interested reader to consult the details there. 

Fourteen experiments were run using a disk galaxy with a prominent bulge \textrm{with $B/D=0.5$} as primary galaxy (equivalent to an S0b galaxy, using 185K particles in total for the simulation). The other two experiments have an initial primary galaxy with a small bulge \textrm{with $B/D=0.08$} (similar to an S0c, with 415K particles in total for the simulation). A physically-motivated size-mass scaling was used to ensure that the primary-to-satellite density ratios are realistic, forcing both galaxies to obey the Tully-Fisher relation: $L \sim V^\alphaTF$, being $L$ the galaxy luminosity  and $V$ its rotational velocity \citep{1977A&A....54..661T}. Different values were initially considered for the exponents of this relation spanning the range of observations ($\alphaTF=3.0$, 3.5, and 4.0). A higher $\alphaTF$ exponent implies a denser satellite compared to the central density of the primary galaxy. Satellites are scaled replicas of the primary galaxy model with a big bulge in all the experiments. 

The galaxy models were built using the {\tt GalactICS} code \citep{1995MNRAS.277.1341K}. The disks of the primary galaxies and the satellites follow exponential surface density profiles radially and vertically. Satellites and primary galaxies were allowed to relax in isolation for about ten disk dynamical times prior to placing them in orbit for the merger simulations, to assure that no relevant resonant structures or spiral patterns appear in any disks. This guarantees that our initial galaxies are S0's and that the evolution observed during the merger is basically driven by it, even the transitory bars and spirals that appear, and not by intrinsic  disk instabilities.

The primary galaxy with a large bulge scales to the Milky Way (MW) considering $R=4.5$\,kpc, $v=220$\,km s$^{-1}$, and $M=5.1\times 10^{10} M_\odot$ as the units of length, velocity, and mass, respectively (the corresponding time unit then being 20.5\,Myr). The primary galaxy with a small bulge matches NGC\,253 using as units $R=6.8$\,kpc, $v=510$\,km s$^{-1}$, and $M=2.6\times 10^{11} M_\odot$ instead (in this case, the time unit is 11.7\,Myr). These values, especially when using an appropriate $M_\mathrm{lum}/L$ ratio, yield total mass-to-light ratios close to observations in both galaxy models ($M/L\sim 10$, where $M = M_\mathrm{lum}+M_\mathrm{dark}$). 

The galaxies are set in parabolic orbits with initial separation equal to 15 times the primary disk scale-length in all the experiments. Two pericentre distances have been analysed: a short one (equal to $h_\mathrm{D}$) and a long one (equal to $8h_\mathrm{D}$). The initial inclinations between the orbital plane and the primary galactic plane were fixed to 30\deg\ in direct orbits and to 150\deg\ in retrograde orbits. The evolution of the models was computed using the {\small TREECODE\/} \citep{1987ApJS...64..715H,1989ApJS...70..419H,1990ApJ...356..359H,1990ApJ...349..562H} and the GADGET-2 codes \citep{2001NewA....6...79S,2005MNRAS.364.1105S}. Using a softening length of $\varepsilon=0.02$ in model units and an opening angle of $\theta = 0.6$, the codes compute forces to within 1\% of those given by a direct summation and preserve the total energy to better than 0.1\%. We evolved all models for $\sim$2-4 halo crossing times beyond a full merger to allow the final remnants to reach a quasi-equilibrium state with a good conservation of energy in all runs. The time period between the full merger and the end of each simulation ranges from 0.4 to 1.1 Gyr depending on the model, for the scalings commented above (see Table\,2 in EM06 and Table\,3 in EM11 for the full-merger and total-run times of each model). 

Table\,\ref{tab:models} lists the main characteristics of each merger experiment. Encounters with galaxy luminous mass ratios between the satellite and the primary galaxy equal to 1:6, 1:9, and 1:18 have been run. We refer to each model throughout the paper according to the code used in EM11: M$m$P[l/s][D/R][b/s], where $m$ indicates the bulge-to-satellite mass ratio ($m=6$, 9, or 18 for models with luminous mass ratios equal to 1:6, 1:9, 1:18, respectively), "Pl" indicates long pericentre and "Ps" short pericentre, "D" or "R" describes the spin-orbit coupling ("D" for direct encounters and "R" for retrograde ones), and the next "b" or "s" letter indicates if the primary galaxy had a big or small bulge. We append to the code a final "TF3" or "TF4" suffix to include the models from the EM06 sample that assume $\alphaTF=3.0$ or 4.0, respectively (excluded in EM11). Otherwise, a scaling with $\alphaTF=3.5$ must be assumed. For more details, see EM06 and EM11.

\begin{sidewaysfigure*}[!]
\begin{center}
\vspace{18cm}
\includegraphics*[width=0.8\textwidth,angle=0]{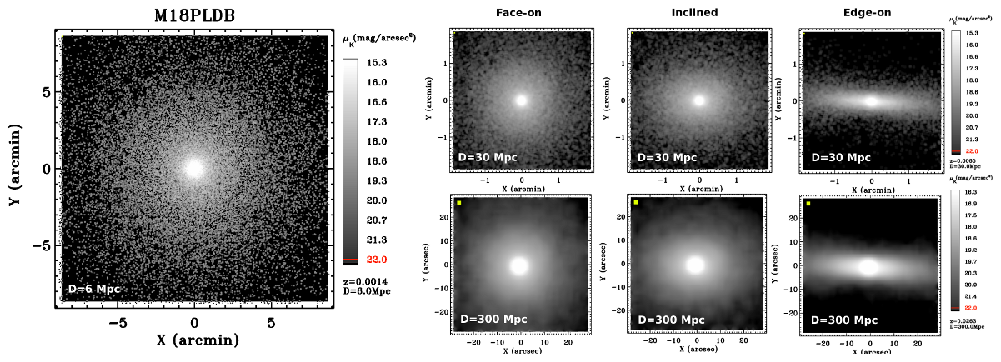}
\includegraphics*[width=0.8\textwidth,angle=0]{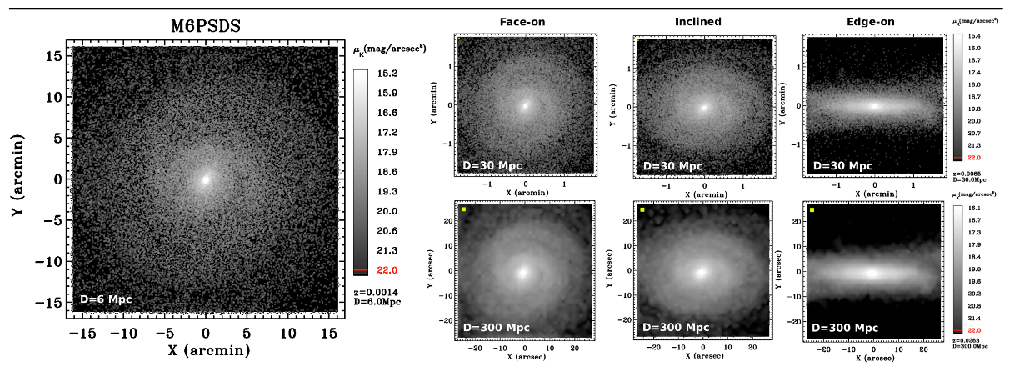}
\caption{Simulated $K$-band images of the final remnants of models M18PlDb (upper figure, experiment k in Table\,\ref{tab:models}) and M6PsDs (lower figure, experiment e in the same Table). We show different galaxy views for the two further simulated distances. The largest frame on the left of each figure corresponds to a face-on view of the remnant at $D=6$\,Mpc. Surface brightness units in the $K$ band are in mag\,arcsec$^{-2}$. A limiting magnitude of $\mu _K=22$\,mag\,arcsec$^{-2}$ and a spatial resolution of FWHM$=0.7$\,arcsec have been assumed (see the text for more details). A grey colour palette is used to represent the different surface brightness levels, set according to a logarithmic scale to enhance spiral patterns in the images with respect to the central bulge. Whiter colours indicate brighter regions (consult the levels bars). The size of each frame fits exactly the apparent angular size of the galaxy for the distance considered in each case. There are yellow squares at the top left corner of each frame indicating the size of the resolution element in each image (0.7$\times$0.7\,arcsec$^2$). }\label{fig:mosaics}
\end{center}
\end{sidewaysfigure*}

\begin{table*}[t!]
\centering 
\caption{Bulge S\'ersic indices, bulge-to-disk luminosity, and scale-length ratios of the original primary galaxies and final remnants of the merger experiments}
\vspace{0.1cm}
\label{tab:photometricparameters}
{\small
\centering
\begin{tabular}{lccc}
\hline\hline
\multicolumn{1}{c}{Model} & \multicolumn{1}{c}{$n$} & \multicolumn{1}{c}{$B/D$} &  \multicolumn{1}{c}{$r_\mathrm{e} / h_\mathrm{D}$} \\
\multicolumn{1}{c}{(1)} & \multicolumn{1}{c}{(2)} & \multicolumn{1}{c}{(3)} & \multicolumn{1}{c}{(4)}\\\hline\vspace{-0.3cm}\\
Original galaxy, $B/D=0.5$ & 0.92$\pm$0.12 &  0.51$\pm$0.03     &  0.190$\pm$0.006 \\
Original galaxy, $B/D=0.08$ & 0.52$\pm$0.02 &  0.0760$\pm$0.0024 &  0.1611$\pm$0.0005 
\vspace{0.05cm}\\\hline\vspace{-0.3cm}\\
(a) M6 Ps Db & 1.7$\pm$0.4   &  0.82$\pm$0.10     &  0.164$\pm$0.005 \\
(a2) M6 Ps Db TF3 &  1.6$\pm$0.3   &  0.77$\pm$0.08     &  0.17$\pm$0.01 \\
(a3) M6 Ps Db TF4 & 1.8$\pm$0.6   &  0.9$\pm$0.3       &  0.169$\pm$0.011 \\
(b) M6 Ps Rb & 1.7$\pm$0.4   &  0.70$\pm$0.10     &  0.182$\pm$0.008 \\
(c) M6 Pl Db & 1.11$\pm$0.04 &  0.676$\pm$0.024   &  0.2396$\pm$0.0002 \\
(d) M6 Pl Rb & 1.02$\pm$0.06 &  0.55$\pm$0.03     &  0.2734$\pm$0.0014 \vspace{0.1cm}\\
(e) M6 Ps Ds &  2.29$\pm$0.16 &  0.179$\pm$0.012   &  0.160$\pm$0.008 \\
(f) M6 Ps Rs  & 2.44$\pm$0.22 &  0.140$\pm$0.012   &  0.143$\pm$0.010 
\vspace{0.1cm}\\\hline\vspace{-0.3cm}\\
(g) M9 Ps Db      & 1.5$\pm$0.3   &  0.770$\pm$0.08    &  0.160$\pm$0.005 \\
(g2) M9 Ps Db TF3 & 1.3$\pm$0.3   &  0.65$\pm$0.06     &  0.166$\pm$0.009 \\
(g3) M9 Ps Db TF4 &  1.3$\pm$0.3   &  0.75$\pm$0.08     &  0.174$\pm$0.004 \\
(h) M9 Ps Rb & 1.9$\pm$0.8   &  0.72$\pm$0.18     &  0.165$\pm$0.015 
\vspace{0.05cm}\\\hline\vspace{-0.3cm}\\
(i) M18 Ps Db & 1.5$\pm$0.3   &  0.60$\pm$0.08     &  0.157$\pm$0.013 \\
(j) M18 Ps Rb & 1.6$\pm$0.3   &  0.68$\pm$0.07     &  0.159$\pm$0.006 \\
(k) M18 Pl Db & 1.05$\pm$0.05 &  0.60$\pm$0.03     &  0.2206$\pm$0.0002 \\
(l) M18 Pl Rb & 1.06$\pm$0.05 &  0.51$\pm$0.03     &  0.1974$\pm$0.0004  \\\hline\\
\end{tabular}
\begin{minipage}[t]{\textwidth}\vspace{-0.2cm}{\small
\emph{Columns}: (1) Model code from Table\,\ref{tab:models}, (2) bulge S\'ersic index, (3) bulge-to-disk mass ratio derived from the S\'ersic$+$exponential fit (equivalent to bulge-to-disk luminosity ratio in the $K$-band, see Sect.\,\ref{Sec:imageSimulations}), (4) bulge-to-disk scale-lengths ratio (defined as the ratio of the bulge effective radius to the disk exponential scale-length).}
\end{minipage}
}
\end{table*}

\section{Image simulations of the remnants}
\label{Sec:imageSimulations}

We have analysed the morphology of our remnants to check whether they can be considered as typical S0's or not (e.g., ellipticals or extremely distorted systems instead). Despite the definition of very efficient quantitative morphological indices \citep{2008ApJ...672..177L}, visual inspection is still the most trustworthy method to classify galaxies morphologically \citep{2007MNRAS.382.1415S,2009ApJ...697.1971J,2010MNRAS.401.1043D}. So, we have classified visually the morphology of the remnants using artificial images simulating the limiting magnitudes and spatial resolutions of specialized surveys. In particular, we have used the NIRS0S sample as a reference \citep[Near-IR S0 galaxy Survey, see][]{2011MNRAS.418.1452L}. As our models trace the dynamical evolution of the stellar mass, we have considered a mass-to-light ratio in the $K$-band of $M_\mathrm{stars}/L_\mathrm{K}\sim 1$\, $M_{\odot}/L_{\odot,\mathrm{K}}$, which is typical of early-type galaxies \citep{2000Ap.....43..145R,2004MNRAS.347..691P}. Therefore, a direct equivalence between the surface density and the $K$-band surface brightness maps can be established in our models for a given mass scaling. 

We have assumed the mass and length scaling commented in Sect.\,\ref{Sec:models}. The primary galaxy in the models with big primary bulge is scaled to the mass and size of the MW, while the one with small bulge is scaled to NGC\,253. Note that the young stellar populations or dust content of these two real galaxies (which are spirals) are not expected to affect significantly to the stellar mass and typical $K$-band scale-lengths of these galaxies. We have performed estimates of the $K$-band magnitude of our  remnants assuming typical star formation histories for spiral and quiescent galaxies. We find that the difference in the total $K$-band magnitude of the galaxy between assuming quiescent or exponentially-decaying star formation histories typical of Sc and Sb galaxies is 0.06 mag at most. Therefore, if the galaxies to which we scale were quiescent, their luminosity and size in this band would be practically the same. This simple exercise guarantees a realistic scaling of our models in this band.

For the MW, a radius $R=13.9$\,kpc \citep{2011ApJ...733L..43M} and a total magnitude $K_\mathrm{T}=-23.8$\,mag have been adopted, both measured in the $K$ band. This last value has been estimated through the conversion of the MW $B$-band total magnitude  \citep[$B_\mathrm{T}=-20.3$\,mag, see][]{1997A&A...326..897H} to the $K$ band, assuming the typical colour of nearby early-type galaxies in the 2MASS Large Galaxy Atlas \citep[$B-K \sim 4$\,mag, ][]{2003AJ....125..525J}. For NGC\,253, we consider a radius $R=14.2$\,kpc and a total magnitude $K_\mathrm{T}=-23.8$\,mag, also in the $K$-band. The absolute $K$-band magnitude of this galaxy is derived assuming the apparent $K$-band magnitude and the distance to this galaxy reported by \citet[$m_K=3.77$\,mag at $D=3.5$\,Mpc]{2003AJ....125..525J}. 

As indicated above, we have considered the limiting $K$-band magnitude and spatial resolution of NIRS0S in our simulated images. This sample consists of $\sim 180$ early-type nearby galaxy disks ($\sim 120$ being S0's) with visual morphological classifications from the Third Reference Catalogue \citep[RC3, see][]{1991trcb.book.....D,1994AJ....108.2128C}. This survey has obtained multi-component decomposition of these galaxies in the $K$-band, to ensure that the derived photometric parameters are not affected severely by dust and star formation. The limiting magnitude of the survey is $\mu_K\sim 21$-22\,mag arcsec$^{-2}$, and the $K$-band images have an average seeing with FWHM$\sim 0.7$\,arcsec. The simulated images have been convoluted by a Gaussian with this FWHM value to simulate the blurring effect associated to any PSF, and any structure below this limiting magnitude has been removed in the images. The typical total $K$-band absolute magnitude of the NIRS0S galaxies is $K_\mathrm{T}\sim -24$\,mag, a very similar value to those exhibited by the two galaxies selected for scaling our models. 

We have considered three characteristic distances for placing our remnants: the minimum, average, and maximum distances of the galaxies in the NIRS0S sample ($D\sim 6$, 30, and 300\,Mpc). The angular scale corresponding to each distance is estimated assuming a concordant cosmology \citep[][]{2007ApJS..170..377S}. Both the cosmological dimming and the K-corrections at first order of approximation are included in the image simulation, albeit their effects are insignificant for such nearby systems. The morphology of the remnants at three different inclinations with respect to the line of sight have been analysed (face-on, inclined by $\sim 45^\mathrm{o}$, and edge-on). We have assumed that the typical noise of the $K$-band images of NIRS0S is negligible compared to the signal of the images relevant for morphological classification, and thus, no noise has been included in the images. We have checked that the surface brightness is conserved with distance. 

Figure\,\ref{fig:mosaics} shows the resulting simulated images for the remnants of models M18PlDb and M6PsDs for different galaxy inclinations and at the three distances considered ($D=6$, 30, and 300\,Mpc). The remnants exhibit a relaxed bulge$+$disk structure with none or weak spiral arms at all distances and inclinations under consideration. This occurs in all the simulated remnants, meaning that all our dry intermediate and minor mergers give rise to S0-like systems, relaxed enough to be considered as undisturbed (the time passed since the full merger varies from $\sim 0.4$ to $\sim 1.1$\,Gyr, depending on the model, see Sect.\,\ref{Sec:models}). 

No bars are observed in any models with big primary bulges, but those with small primary bulges present weak bar-like distortions and residual spiral patterns in the centre (see the figure). This difference arises in the lower initial $B/D$ ratio of the primary galaxy, as the presence of any central mass concentration tends to stabilize the disk \citep[][]{1990ApJ...363..391P,1999ApJ...510..125S,2002A&A...392...83B,2005MNRAS.363..496A,2005ApJ...628..678D,2005MNRAS.357..753G,2008MNRAS.384..386C}. This makes spiral arms and bar distortions to be more efficiently inhibited in the remnants of models with initial big primary bulges than in those with small ones. Nevertheless, these inner structures are insignificant compared to the bulge and disk component, so we can assume that the luminous mass in our remnants is distributed basically according to an axi-symmetric bulge$+$disk structure. 

\section{Bulge-disk photometric decompositions of the remnants} 
\label{Sec:photometricParameters}

We have derived the bulge-disk photometric parameters from azimuthally-averaged surface brightness profiles for the luminous matter, accounting for the axial symmetry of all the remnants (see Sect.\,\ref{Sec:imageSimulations}). In general, the global photometric parameters derived from 1D and 2D fits in axi-symmetric cases are quite similar within errors \citep{1996A&A...313...45D,1998PhDT........29M}. The 1D profiles present a much better signal-to-noise as compared to the 2D surface brightness maps, ensuring a higher accuracy of the fits \citep{2010AJ....139.2097P}. We have derived the 1D surface brightness profiles with all the merger remnants viewed face-on to compare with observations, which usually correct their results for the effects of galaxy inclination. 

We have fitted a combined S\'{e}rsic$+$exponential function to the density profiles, as in EM06. The S\'ersic profile has been extensively used for modelling bulge and elliptical surface brightness profiles \citep{1968adga.book.....S}:

\begin{equation}
I(r)=I_{\mathrm{e}} \exp\,\{- b_{n}\, [ ( r/r_{\mathrm{e}}) ^{1/n}-1] \} , \label{Eq:Sersic}
\end{equation}

\noindent where $r_{\mathrm{e}}$ is the bulge effective radius, $I_{\mathrm{e}}$ is the surface brightness at $r_{\mathrm{e}}$, and $n$ is the S\'{e}rsic index, which is an indicator of the bulge concentration. The factor $b_{n}$ is a function of $n$, which may be approximated by $b_{n}=1.9992\, n-0.3271$ in the range $1<n<10$, with an error $<0.15$\% in the whole range \citep{1991A&A...249...99C,2001MNRAS.326..543G}. 

The exponential law adequately describes the radial profiles of most disks \citep[][]{1970ApJ...160..811F}:
\begin{equation}
I(r)=I_{\mathrm{0}} \exp\left( -r/h_{\mathrm{D}}\right) , \label{Eq:disc}
\end{equation}

\noindent $h_{\mathrm{D}}$ being the scale-length of the profile and $I_{\mathrm{0}}$ representing the extrapolated central surface brightness. 

The code used for fitting the radial surface density profiles is described in EM06. Residuals are below $\sim$0.2 mag along the whole radial range, a quite reasonable result compared to typical observational errors. Errors of the photometric parameters are estimated with the bootstrap method \citep{Efron93,Press94}. For all our experiments, we recover a two component system with the outer parts well fitted by an exponential profile (\ie, a disk), while the inner parts are well represented by S\'ersic functions with $1< n < 2.5$ (see Fig.\,3 in EM06 for some examples). Final values of $n$, $B/D$, and $r_\mathrm{e}/h_\mathrm{D}$ derived from these fits to the surface density profiles of the remnants are listed in Table~\ref{tab:photometricparameters}. Our models are scalable in size, because they are collision-less. This means that the bulge effective radius of each remnant can be scaled to any physical size, just considering that the disk scale-length is related to this value according to the $r_\mathrm{e}/h_\mathrm{D}$ ratio of this remnant tabulated in Table\,\ref{tab:photometricparameters}.

\begin{figure}[t!]
\center
\includegraphics[width = 0.5\textwidth, bb= 30 0 481 481, clip]{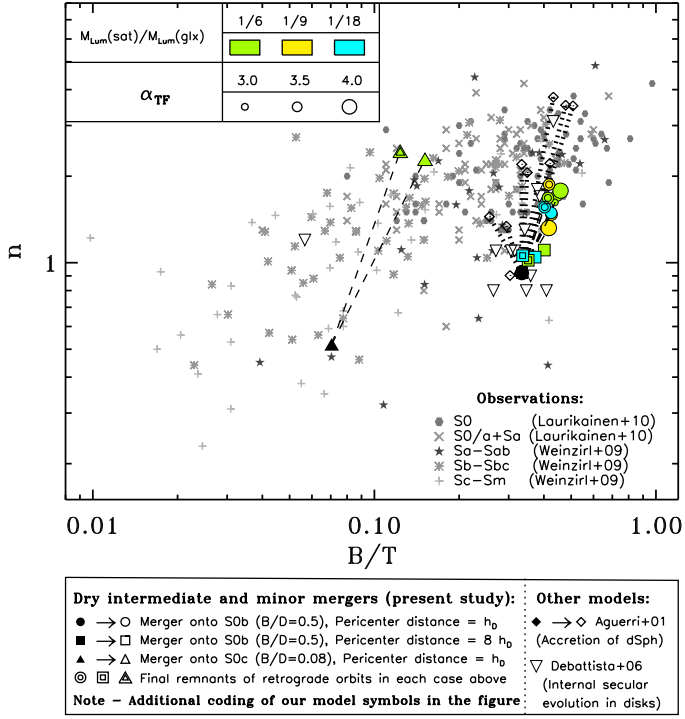} 
\caption{Growth vectors in the log\,($n$)-log\,($B/T$) plane driven by our merger experiments, compared to the observational distributions of S0's and spirals. Each segment starts at the location of the original galaxy model and ends at the final $n$ and $B/T$ values of the remnants. The growth vectors corresponding to the merger models with dense spheroidal satellites by \citet{2001A&A...367..428A} are represented too, as well as the bulges resulting from the internal secular evolution models in pure exponential disks by \citet{2006ApJ...645..209D}. Consult the legend in the figure.}\label{fig:nbt}
\end{figure}

\begin{figure}[!]
\includegraphics[width = 0.5\textwidth, bb= 10 98 481 481, clip]{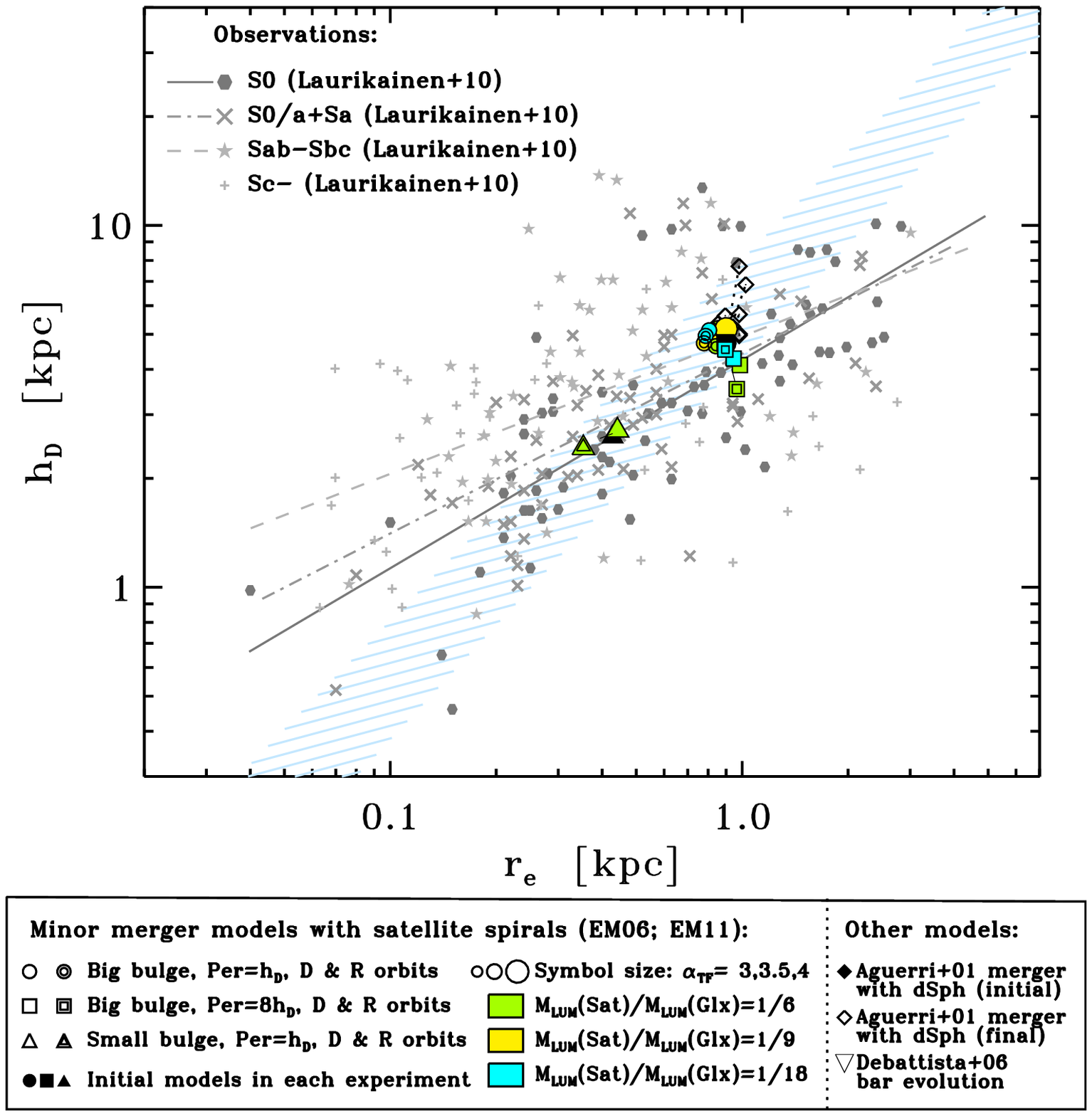} 
\caption{Growth vectors in the log\,($h_\mathrm{D}$)-log\,($r_\mathrm{e}$) plane driven by our merger experiments, compared to the observed distributions of S0's and spirals. The straight lines show the correlations found by L10 for different morphological types  (\emph{red solid line}: for the S0's, \emph{blue dotted-dashed line}: for the S0/a$+$Sa's, \emph{green dashed line}: for the Sab-Sbc's). The blue shaded region indicates the locations in the plane that are covered by our models adjusting the scaling (see Sect.\,\ref{Sec:models}). Note that the change in both scale-lengths is negligible, making the remnants to basically overlap with the original location of the primary galaxies in the plane. Consult the legend for the observational data in the figure. The legend for the models is the same as in Fig.\,\ref{fig:nbt}. } \label{fig:hre}
\end{figure}

\section{Results}
\label{Sec:results}

We have analysed the evolution induced by the mergers in the bulge and disk photometric parameters of the primary galaxy, to compare the final values of $B/D$, $r_\mathrm{e}/h_\mathrm{D}$, and $n$ in the remnants with the observational distributions of these parameters in real spirals and S0's reported by \citet[][]{2009ApJ...696..411W} and L10. These studies use images in NIR bands ($H$ and $K$, respectively). We also compare our results with the evolution entailed by accretion experiments onto an S0b of high-density dSph satellites \citep{2001A&A...367..428A} and with internal secular models of pure exponential disks \citep{2006ApJ...645..209D}.

\subsection{Evolution along the sequence of S0 Hubble types}
\label{Sec:BTn}

The initial $B/D$ ratios of the primary galaxies increase after the merger by a factor $\sim 1.1$-1.8 in the case of big primary bulges and $\sim 1.8$-2.3 in the models with small ones (see Table\,\ref{tab:photometricparameters}). According to the observed correlations between $n$ and $B/D$ with the morphological type \citep[][]{2001AJ....121..820G,2009ApJ...696..411W}, this means that the remnants of the experiments starting with an S0b primary galaxy have become S0a's, while the S0c primary galaxies are transformed into S0b's. 

The dry intermediate and minor mergers lead to a more noticeable growth of the bulge-to-total luminosity ratio ($B/T$) in the experiments with small primary bulges ($\sim 150$\% with respect to the original value) than in those with big primary bulges ($\sim 60$\%), for similar mass ratios and orbits of the encounter (consult Table\,\ref{tab:photometricparameters}). 

Figure\,\ref{fig:nbt} shows the final distributions of all remnants in the log\,($n$)-log\,($B/T$) plane, compared to the observed distributions of real galaxies with different morphological types. For each merger simulation, we plot the initial and post-merger values. The figure shows that dry intermediate and minor mergers onto S0's increase the mass ($B/T$) and concentration of the bulge ($n$), as already reported by previous studies \citep[][]{2001A&A...367..428A,2007ApJ...670..269Y}. Although the initial S0 models are offseted from the location of typical S0's in this plane (the original $n$ values are lower than it should be for the $B/T$ ratio), the remnants are located in the region populated by S0's. The evolution induced by these mergers along the sequence of S0 Hubble types is coherent with the observational fact that the majority of S0's have exponential-like bulges ($1<n<2$), instead of $n>3$ bulges \citep{2005MNRAS.362.1319L,2006AJ....132.2634L,2007ApJ...665.1104B,2009ApJ...696..411W}. The bulges resulting from internal secular evolution as modelled by \citet{2006ApJ...645..209D} overlap in the plane with our remnants.

Figure\,\ref{fig:nbt} shows some trends of the bulge growth driven by the merger with the initial conditions of the model:

\begin{enumerate}
\item The accretion of a satellite by a galaxy with an original big bulge leads to a smoother growth in both $n$ and $B/T$ than in the case of being accreted by a galaxy with a small bulge, relatively to their initial values.

\item Moreover, the minor mergers undergone in orbits with long pericentres produce a much more moderate growth in both $n$ and $B/T$ than analogous mergers with short impact parameters. 

\item Curiously, the evolution in the $n$-$B/T$ plane seems to depend more significantly on the orbit than on the mass ratio of the encounter for the same original galaxy, contrary to previous findings in major merger simulations \citep{2004A&A...418L..27B}.

\item Finally, as already reported in EM06, mergers with denser satellites (i.e., higher $\alphaTF$) lead to more noticeable growths in $n$ than those with low-density satellites.  The high-density satellites of the \citet{2001A&A...367..428A}~models produce more extreme evolution in this plane than the one produced by our models for the same original galaxy. We expect satellites with lower central densities than ours to disrupt completely before reaching the galaxy and contribute to the stellar halo of the galaxy, as already observed in other simulations \citep{2010MNRAS.406..744C}.

\end{enumerate}

Our models suggest that dry intermediate and minor mergers can induce bulge growth in two ways relatively to their initial state, depending on the mass of the original primary bulge.  On the one hand, galaxies with small bulges experience a prominent rise in $n$, and a much more moderate increase of $B/T$. This is because the satellite core ends deposited in the remnant centre without disrupting, rising the mass concentration, and thus $n$ (EM11). On the other hand, galaxies with big bulges experience moderate growths in $n$ and $B/T$  after the accretion of a satellite compared to their original status, with a higher dependence on the orbit than on the mass ratio. This is because satellites disrupt completely before reaching the centre, depositing their mass at longer distances from it (EM06; EM11). 

In conclusion, these models confirm that dry minor and intermediate mergers onto S0's can induce evolution in the $n$-$B/T$ plane towards earlier S0 Hubble types.

\begin{figure*}[t]
\center
\includegraphics[width = 0.48\textwidth, bb= 10 100 481 481, clip]{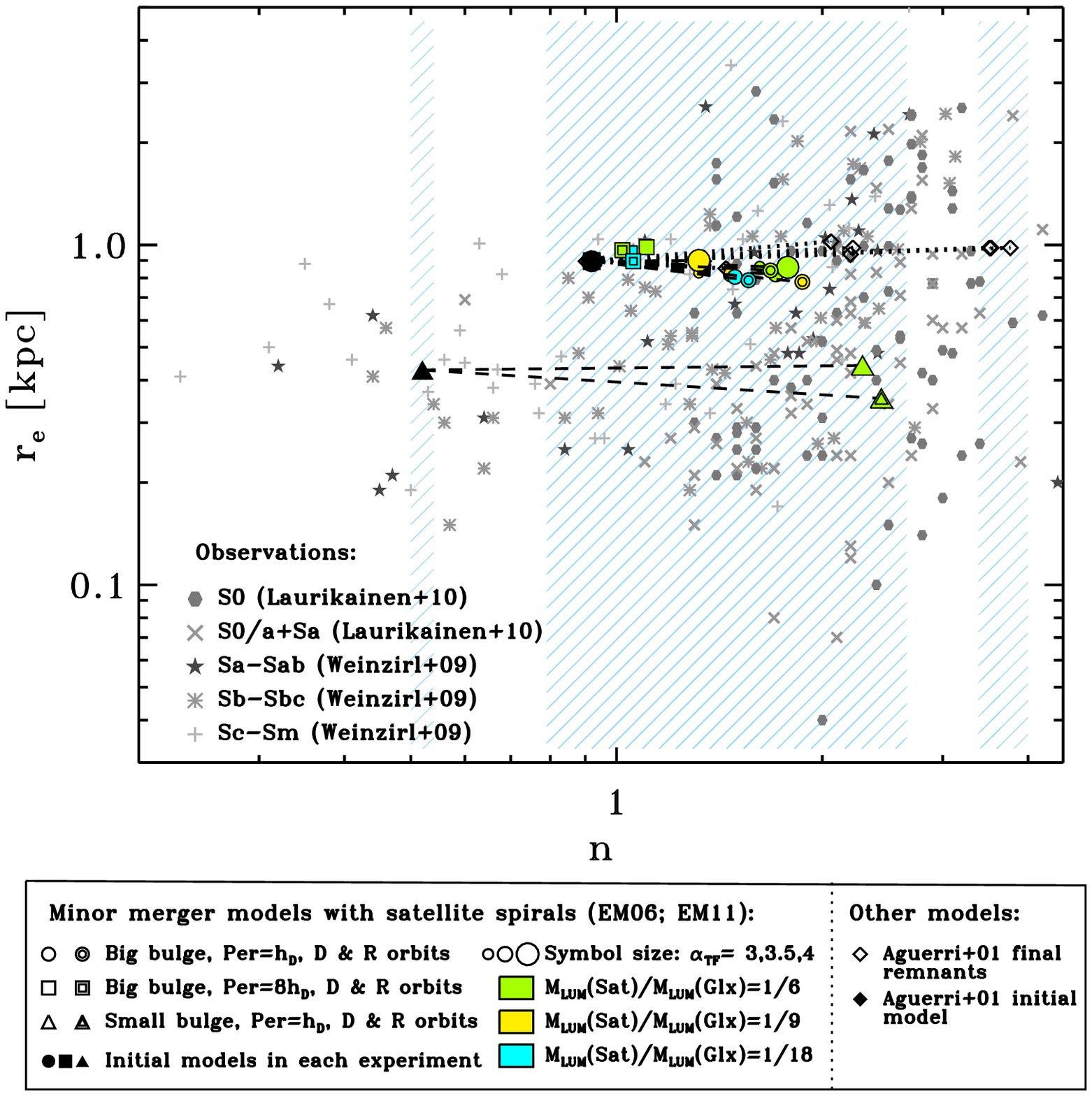} 
\includegraphics[width = 0.48\textwidth, bb= 10 100 481 481, clip]{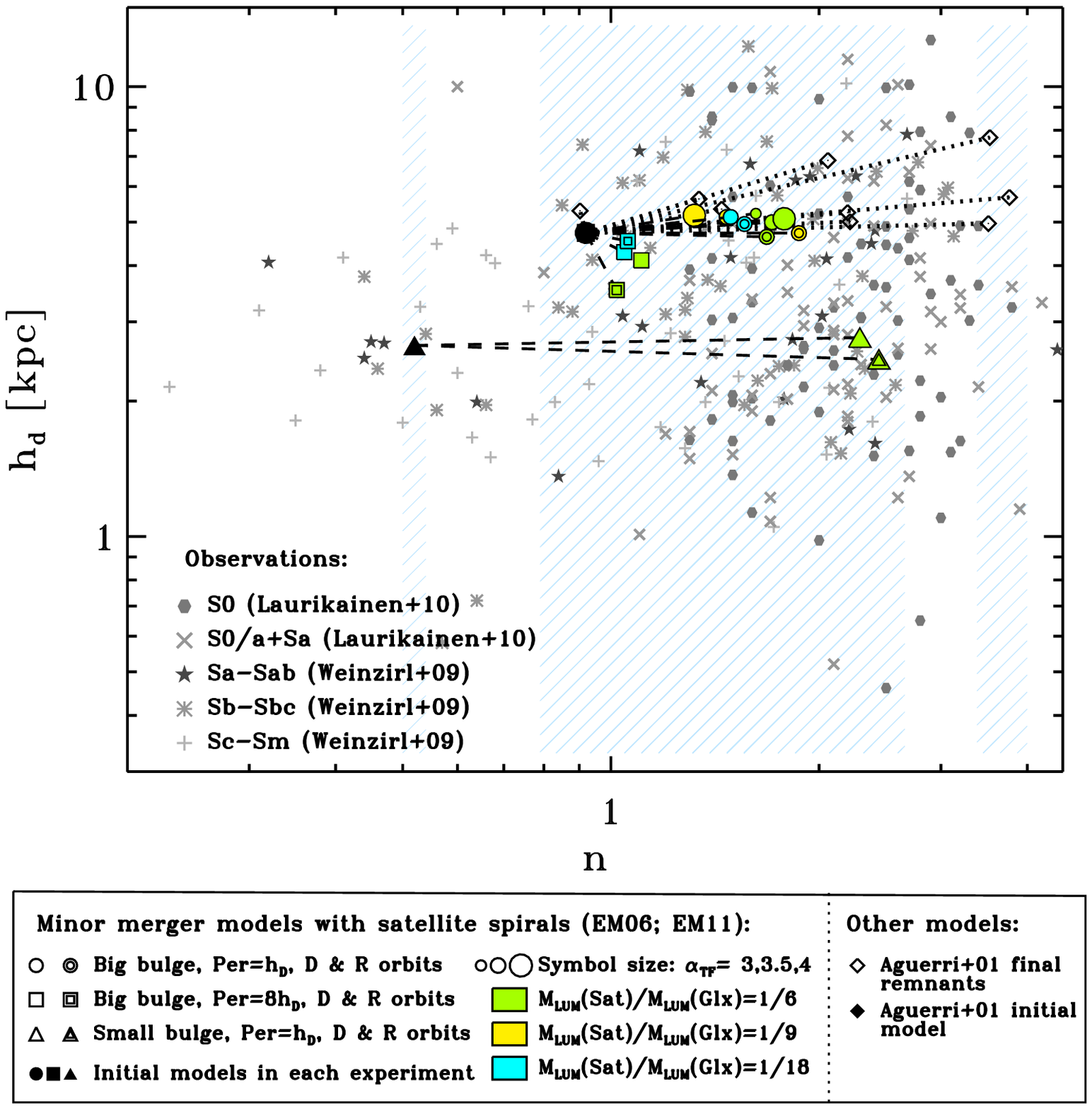} 
\caption{Growth vectors in the log\,($r_\mathrm{e}$)-log\,($n$) and log\,($h_\mathrm{D}$)-log\,($n$) planes driven by our merger experiments, compared to the observational distributions of S0's and spirals. The blue shaded region indicates the locations in the plane that are covered by our models just using a scaling different to the one indicated in Sect.\,\ref{Sec:models}. Consult the legend for the observational data in the figure. The legend for the models is the same as in Fig.\,\ref{fig:nbt}.} \label{fig:rer0n}
\end{figure*}

\begin{figure*}[t]
\center
\includegraphics[width = 0.48\textwidth, bb= 10 100 481 481, clip]{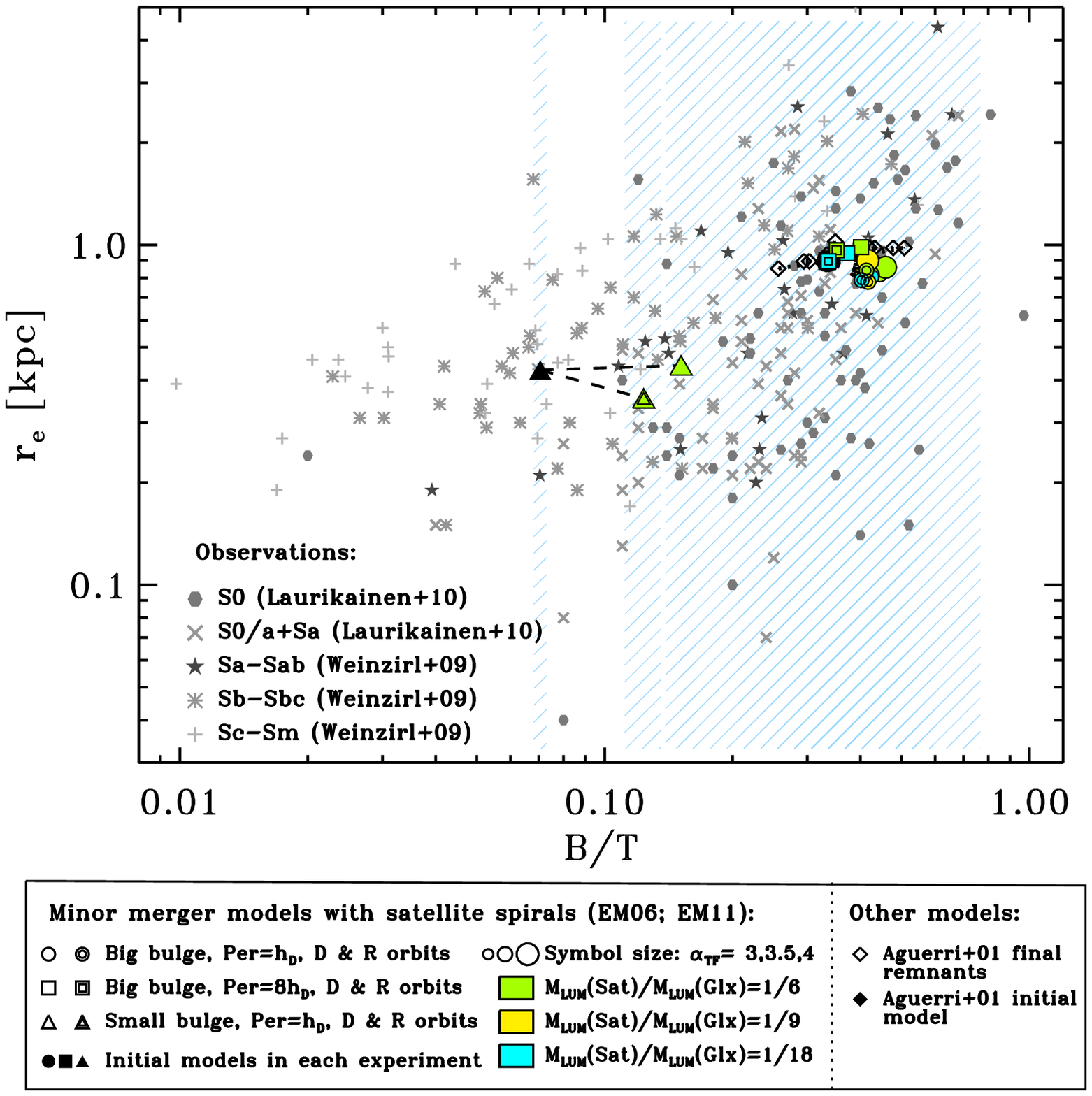} 
\includegraphics[width = 0.48\textwidth, bb= 10 100 481 481, clip]{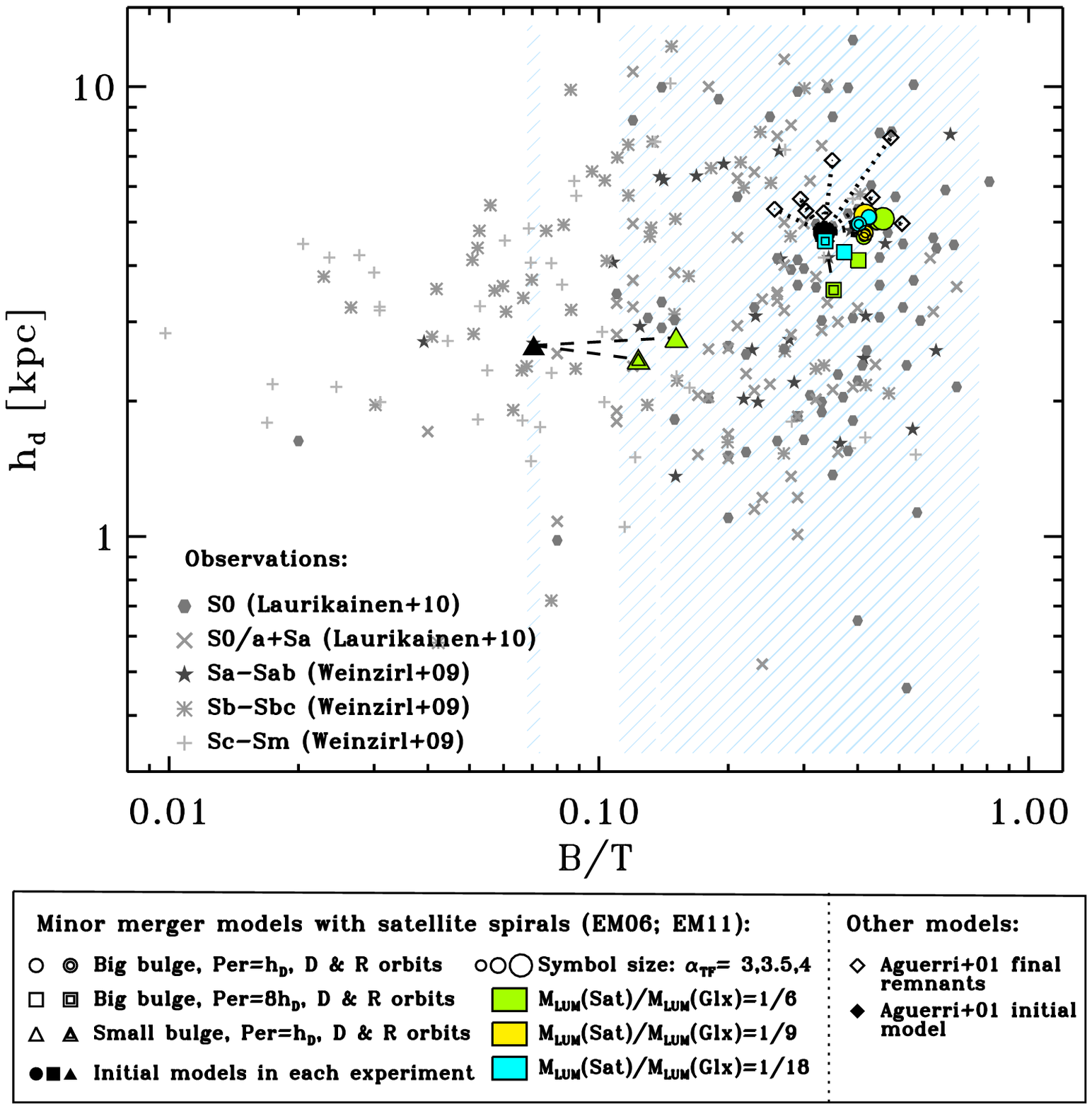} 
\caption{Growth vectors in the log\,($r_\mathrm{e}$)-log\,($B/T$) and log\,($h_\mathrm{D}$)-log\,($B/T$) planes driven by our  merger experiments, compared to the observational distributions of S0's and spirals. The blue shaded region indicates the locations in the plane that are covered by our models just using a scaling different to the one indicated in Sect.\,\ref{Sec:models}. Consult the legend for the observational data in the figure. The legend for the models is the same as in Fig.\,\ref{fig:nbt}.} \label{fig:rer0bt}
\end{figure*}

\subsection{Correlations between the bulge and disk scale-lengths}
\label{Sec:rehn}

In Fig.\,\ref{fig:hre}, we show the location of the final remnants of our merger models in the log\,($h_\mathrm{D}$)-log\,($r_\mathrm{e}$) plane, compared to the observational distributions of real galaxies with different morphological types. We have marked the locations in the plane that would be achievable by our models just assuming a different scaling. The straight lines indicate the correlations in the plane found for different galaxy types by L10. Observationally, galaxies distribute diagonally in this diagram, i.e., the scale-lengths of the bulge and disk components correlate tightly in real galaxies. These correlations are quite similar for different galaxy types (compare the straight lines) and, more relevantly, the distributions of the different types overlap completely in the diagram. The figure shows that the region in the $h_\mathrm{D}$-$r_\mathrm{e}$ plane that can be achieved by our models through different scaling overlaps with the distribution of real S0's basically. 

Our models demonstrate that dry intermediate and minor mergers can produce a negligible change in the location of a galaxy in the $h_\mathrm{D}$-$r_\mathrm{e}$ plane, independently on the $B/T$ of the original S0 and on the properties of the encounter. Therefore, this sort of mergers basically contributes to rise slightly the dispersion of the relation between the scale-lengths of the bulge and disk components of S0's, although they can induce a noticeable bulge growth and move the galaxies towards earlier types. This result indicates that dry intermediate and minor mergers are capable of producing remnants in which the scale-lengths of the bulge and the disk are coupled, if they were already coupled in the original galaxy. Hence, the main conclusion of Fig.\,\ref{fig:hre} is that minor mergers cannot be discarded as feasible mechanisms for the evolution of S0's just on the basis of the strong coupling of the bulge and disk scale-lengths observed in these galaxies.

We plot in Fig.\,\ref{fig:rer0n} the evolution of $r_\mathrm{e}$ and $h_\mathrm{D}$ versus $n$ induced by our minor merger experiments compared to the observational distributions of different morphological types. The blue shaded region indicates the locations of these diagrams that are achievable with our models, just assuming a length scaling different to the one indicated in Sect.\,\ref{Sec:models}. The simulations indicate that the distribution in the plane of S0's could be reproduced through dry intermediate and minor merger events basically, starting from S0's of a late Hubble type. Therefore, Figs.\,\ref{fig:nbt}-\ref{fig:rer0n} prove that the global bulge-to-disk structural properties of real S0's are not incompatible with an evolution driven by gas-poor mergers of low mass ratio. 

Figure\,\ref{fig:rer0bt} shows the evolution of $r_\mathrm{e}$ and $h_\mathrm{D}$ versus $B/T$ induced by our minor merger experiments compared to the observational distributions of different morphological types. Both scale-lengths do not vary by more than $\sim 20$\% of their original value after the merger. The final $B/T$ ratio tends to rise after the mergers, except in a few cases by \citet{2001A&A...367..428A}. The simulations can reproduce the distribution of S0's also in these planes, just assuming a different scaling.

In conclusion, our models demonstrate that the dry intermediate and minor mergers can evolve S0's along the sequence of S0 Hubble types towards earlier types, fulfilling global bulge-to-disk structural relations compatible with observations.

\begin{figure*}[t]
\center
\includegraphics[width = 0.48\textwidth, bb= 0 100 481 481, clip]{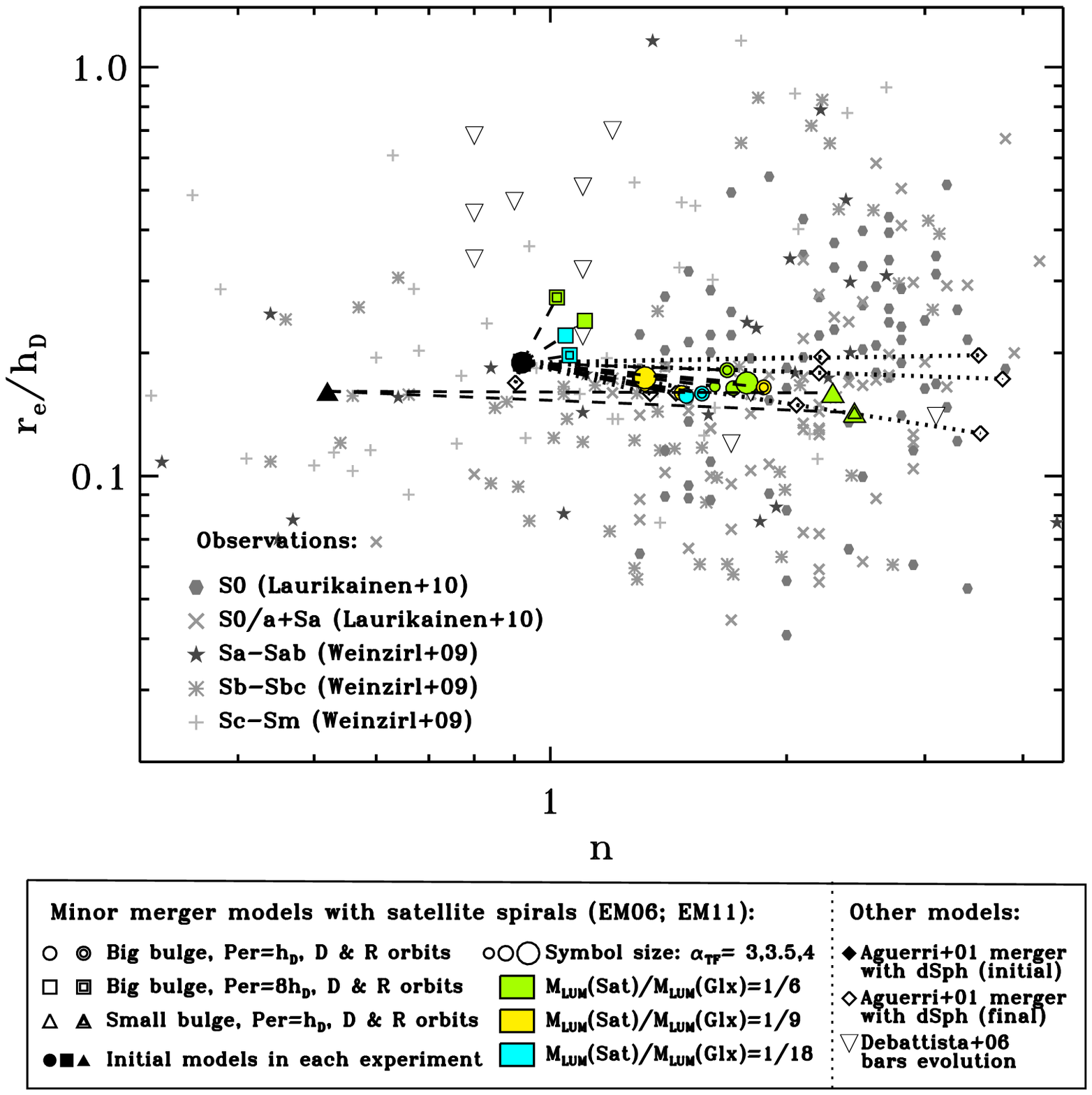} 
\includegraphics[width = 0.48\textwidth, bb= 10 100 481 481, clip]{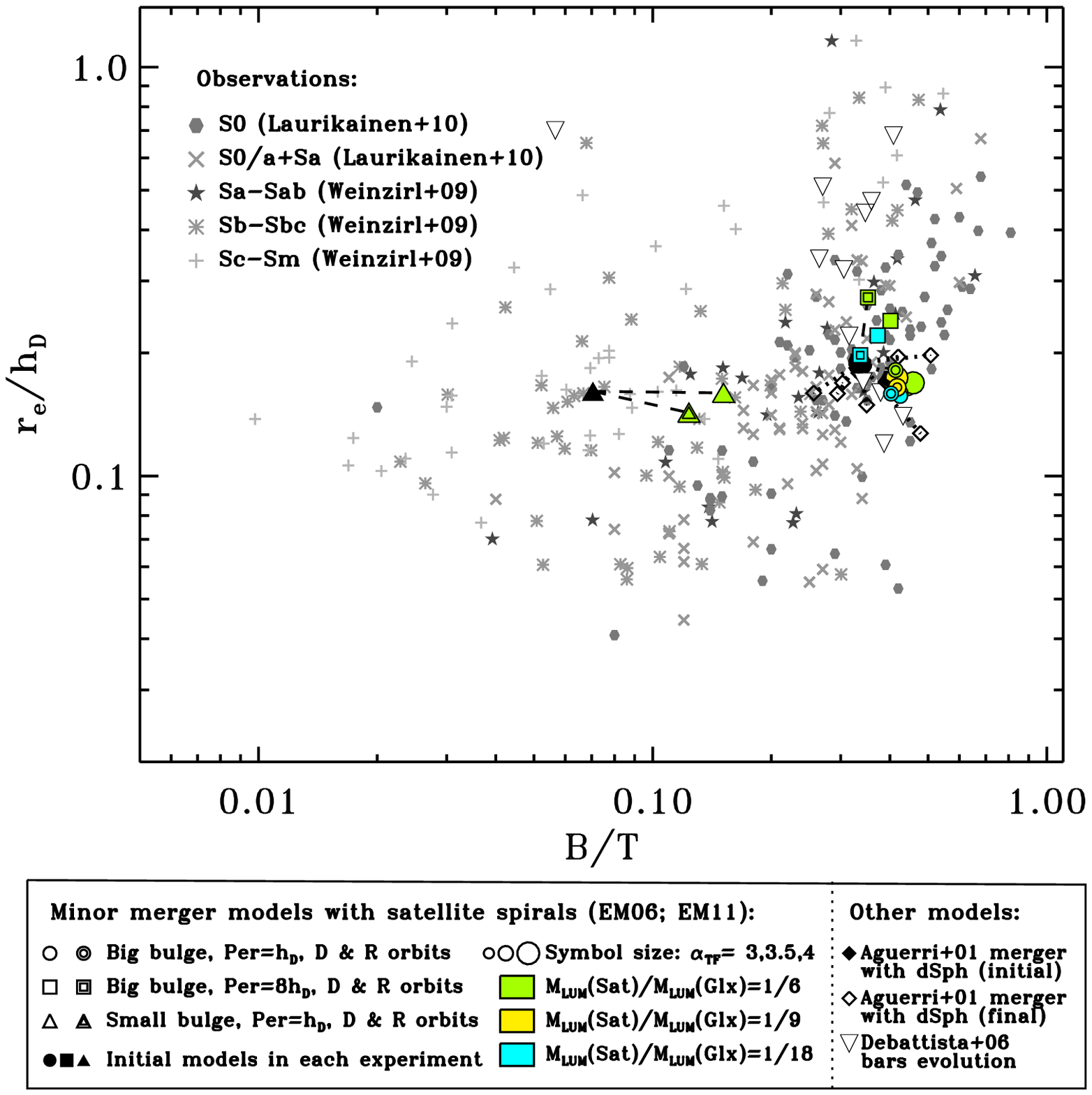} 
\caption{Growth vectors in the log\,($r_\mathrm{e}/h_\mathrm{D}$)-log\,($n$) and log\,($r_\mathrm{e}/h_\mathrm{D}$)-log\,($B/T$) planes driven by our merger experiments (left and right panels, respectively), compared to the observed distributions of S0's and spirals. Consult the legend for the observational data in the figure. The legend for the models is the same as in Fig.\,\ref{fig:nbt}. } \label{fig:rehbt}
\end{figure*}

\subsection{A scale-free Sequence of S0 Hubble types}
\label{Sec:HubbleSequence}

The lack of correlation of $r_\mathrm{e}/h_\mathrm{D}$ with $n$ and $B/T$ for spirals led many authors to conclude that the Hubble sequence is scale-free \citep{1996A&A...313...45D,1996ApJ...457L..73C,2001MNRAS.326..543G,2003ApJ...582..689M,2007ApJ...665.1104B}. Figure\,\ref{fig:rehbt} represents the evolution driven by our minor merger experiments in $r_\mathrm{e}/h_\mathrm{D}$ as a function of the S\'ersic index (left panel) and $B/T$ ratio (right panel), compared to observational data. The data support the previous finding also for S0's: there is a wide dispersion in the $r_\mathrm{e}/h_\mathrm{D}$ ratio for any $n$ and $B/T$ values found in S0's too, although this dispersion rises at the region of high $n$ and $B/T$ values, where S0's tend to accumulate (this is more noticeable in the right panel).

The remnants of our experiments are also compatible with the observed distributions of S0's in these planes. Our models show that dry mergers of low mass ratios onto S0's make them to move towards higher $n$ and $B/T$ values, keeping the original $r_\mathrm{e}/h_\mathrm{D}$ ratio. Although the minor merger can produce a significant rise in $n$ or $B/T$ depending on the initial conditions of the encounter,  the ratio $r_\mathrm{e}/h_\mathrm{D}$ is preserved or slightly dispersed. This demonstrates that the evolution induced by these mergers is compatible with three observational facts: 1) with the accumulation of S0's of early-types towards high $n$ and $B/T$; 2) with the lack of correlation of  $r_\mathrm{e}/h_\mathrm{D}$ with the morphological type of the S0, because this ratio depends exclusively on the original galaxy experiencing the accretion, not on the minor merger itself; and 3) the slight increase of dispersion in $r_\mathrm{e}/h_\mathrm{D}$ as the Hubble type becomes earlier. 

Note that the internal secular evolution models of pure exponential disks by \citet{2006ApJ...645..209D} generate bulges that do not overlap with real S0's in Fig.\,\ref{fig:rehbt}, even those without gas. Their structures are more similar to the bulges of spirals than to those of S0's. 

Therefore, our models suggest that dry mergers of low mass ratios onto S0's can provide one of the feasible explanations for the buildup of a scale-free sequence of Hubble S0 types, starting from S0c's of late types formed through other mechanisms.

\begin{figure*}[t]
\center
\includegraphics[width =  \textwidth]{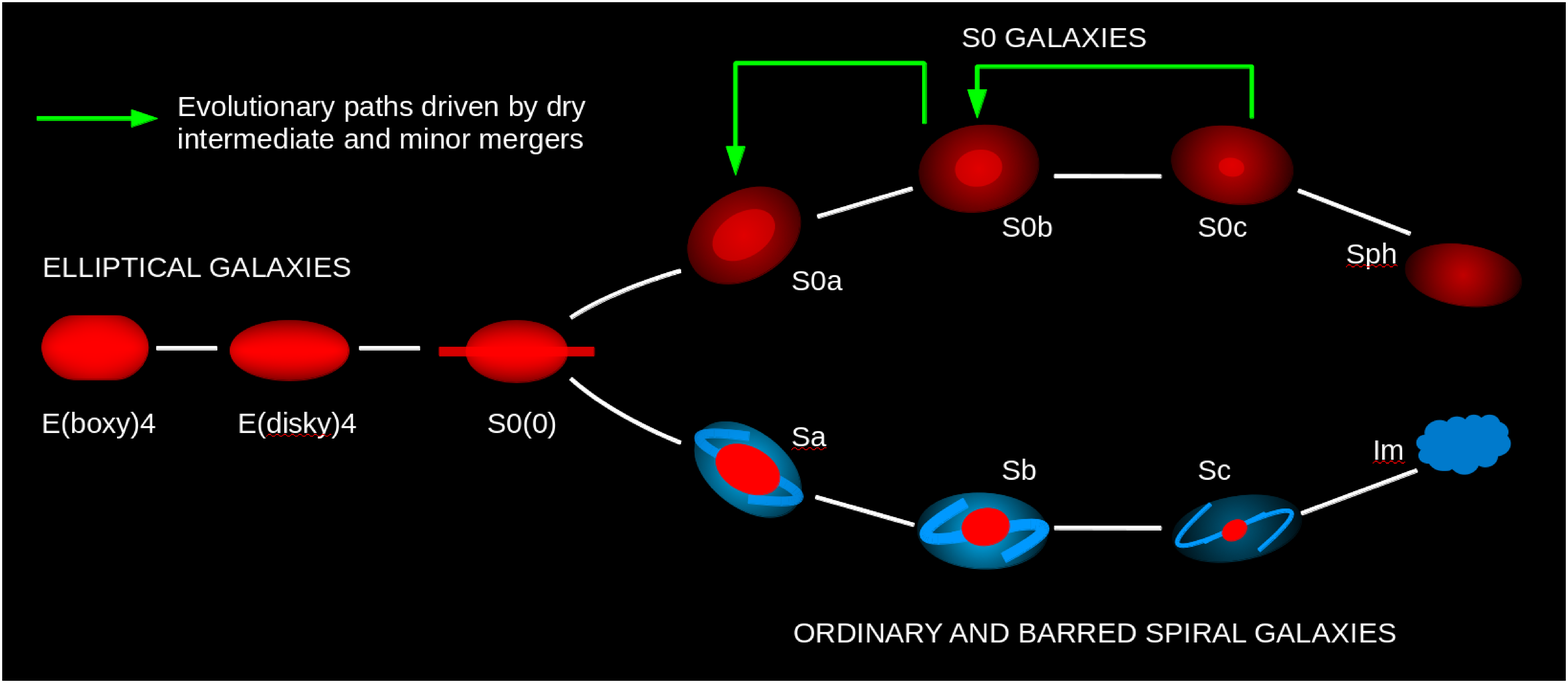} 
\caption{Schematic representation of the evolutionary paths driven by our dry intermediate and minor merger experiments along the sequence of S0 Hubble types. Bulge-to-total luminosity ratios increase towards the left, accordingly to the evolution produced by our simulated mergers. Our models show that minor mergers can induce the evolution of S0b$\longrightarrow$S0a and of S0c$\longrightarrow$S0b, reproducing the global bulge-to-disk structural relations found in real S0's. The revised Hubble tuning fork plotted in the figure is inspired in Fig.\,1 by \citet{2012ApJS..198....2K}. } \label{fig:evolution}
\end{figure*}

\section{Discussion}
\label{Sec:discussion}

Our simulations prove that dry intermediate and minor mergers can make S0 galaxies to evolve towards earlier S0 types, fulfilling  global structural relations observed between the bulges and disks of these galaxies. According to the final $B/T$ ratios of the remnant S0's (see Table\,\ref{tab:photometricparameters}), our experiments show that these mergers induce the following transformations: S0c$\longrightarrow$S0b and S0b$\longrightarrow$S0a. These evolutionary paths are indicated with green arrows in the Hubble Sequence diagram of Fig.\,\ref{fig:evolution}. 

Considering that the growth in $n$ and $B/D$ of successive minor mergers is cumulative \citep[although it seems to saturate at $n\sim 4$-5 values, see][]{2001A&A...367..428A,2007A&A...476.1179B}, it should be possible to reproduce the distributions of S0's in the structural planes shown in Figs.\,\ref{fig:nbt}-\ref{fig:rehbt} through subsequent minor merging. Therefore, the models demonstrate that dry intermediate and minor mergers can give rise to a sequence of S0 Hubble types parallel to the one of spirals, starting from S0c's. 

As commented in Sect.\,\ref{Sec:introduction}, mergers are usually discarded as feasible evolutionary mechanisms for S0's, attending to several observational facts that people interpret as exclusive of pure internal secular evolution. By the contrary, the present models prove this interpretation to be questionable, as many of these observational results can be explained or are compatible with an evolution driven by dry intermediate and minor mergers. 

In particular, the fact that the bulge and disk scale-lengths in S0's are tightly correlated is compatible with an evolution driven through these mergers, because they nearly preserve the original bulge-to-disk scale-lengths ratio. This evolutionary scenario only requires the correlation between the two scale-lengths to have been established at an early epoch of bulge growth. But this is ensured by observational data, as L10 demonstrate that this correlation does not depend on galaxy type. This means that it is completely feasible to partially build up the parallel sequence of S0 Hubble types through dry minor mergers onto S0's, as these processes modify the sizes of both components in such a way that their ratio is nearly conserved, moving galaxies towards earlier Hubble types at the same time. Moreover, these mergers introduce some dispersion in the $r_\mathrm{e}/h_\mathrm{D}$ ratio  compatible with observational data, providing a straightforward explanation for the lack of trend of the $r_\mathrm{e}/h_\mathrm{D}$ with the Hubble type in S0 systems.

Curiously, another observational result compatible with a dry minor merger evolution is the high fraction of pseudobulges found in S0's \citep{2005MNRAS.362.1319L,2006AJ....132.2634L,2007MNRAS.381..401L}. Note that the S0's resulting from our dry minor mergers have $n\lesssim 2$ and $B/T<0.4$. Although the definition of pseudobulges take into account several aspects of the shapes and dynamics of the central regions of galaxies, the bulges obtained in our simulations fulfill some criteria that define a pseudobulge, according to \citet{2004ARA&A..42..603K}. Therefore, dry minor mergers cannot be discarded as possible evolutionary mechanisms of S0's just arguing that many of them exhibit pseudobulges, assuming that these bulge structures are associated exclusively with pure internal secular evolution. Our models prove that dry minor mergers can generate pseudobulges if the original S0 had already one.

In fact, our minor merger experiments show that these events can induce internal secular evolution, even in the absence of gas and star formation (EM06; EM11). In EM06, we reported that the bulge growth in these experiments was basically due to the injection of disk material towards the galaxy centre through the transient disk distortions induced by the interaction. The satellite material deposition was essential for rebuilding an exponential disk at central and intermediate radii in the remnant, but not for the bulge growth. These transitory distortions (such as spirals, bars, and ovals) are usually ascribed to pure internal secular evolution, but these models prove that minor and intermediate mergers can be highly efficient at inducing them, even in gas-poor events.

Additionally, the lower frequency of bars detected in S0's compared to the one found in Sa$+$S0/a's is not well understood in an evolutionary scenario majoritarily driven by pure internal secular processes \citep[][]{1979ApJ...227..714K,2007MNRAS.381..401L,2009ApJ...692L..34L,2010ApJ...721..259B,2010ApJ...714L.260N}. Sa$+$S0/a's are the most probable progenitors of S0's according to this scenario, and thus, their bar fractions should be similar, unless the process entails a bulge growth enough to inhibit bar distortions. But this does not seem to be the case. According to estimates by \citet{2004ARA&A..42..603K}, internal secular evolution can give rise to bulges with $B/T\sim 0.16$ at most, whereas S0's have $B/T\sim 0.25-0.30$ typically \citep{2007ApJ...665.1104B,2007MNRAS.381..401L,2009ApJ...696..411W}. Nevertheless, the difference in bar fractions between these two galaxy types is coherent with an S0 evolution partially driven by dry minor mergers. We have shown that mergers are quite efficient rising the concentration and mass of the central bulge, even for gas-free encounters with mass ratios as low as 1:18. This process tends to suppress bar formation, as commented above, being compatible with a decrease of the bar fraction in S0's compared to any other Hubble type.

The high fraction of dynamically-cold inner components found in S0's is another observational result that is usually put forward against the possible merger-driven evolution in these galaxies  \citep[see references in][]{2012ApJS..198....2K}. Although we have not focussed the present study on this, it is remarkable that our merger models also show this concept to be questionable. In EM11, we show that all our remnants host thin rotationally-supported inner components made out of disrupted satellite material, such as inner disks, inner rings, pseudo-rings, nested inner disks, and central spiral patterns. Consequently, the high number of S0's hosting dynamically-cold stellar components might be pointing to their merger-related origin instead of to the contrary fact, as usually interpreted.

Moreover, the high number of ovals and lenses found in S0's ($\sim 97$\%) is considered as a result pointing to the relevance of bars (and thus, of pure internal secular evolution) in the formation of these systems \citep[][]{2009ApJ...692L..34L}. But, again, this result does not exclude a minor-merger origin for these structures. Many of our remnants have ovals, probably derived from the transitory bar-like distortions induced by the encounter in the original galaxy disk, as we also show in EM11.

Finally, the apparently non-distorted morphology exhibited by an S0 cannot be considered as a sign excluding recent past merging either; first, because very deep imaging may reveal faint tidal features that are not detectable in typical images in many cases \citep[see][]{2011MNRAS.417..863D,2012ApJ...753...43K}, and secondly, because this study proves that remnants of dry intermediate and minor mergers can exhibit a relaxed morphology only 1-2\,Gyr after the complete merger for the typical observing conditions of current surveys.

We must remark that all these results do not exclude the relevant contribution of other processes to the buildup of S0's, but the present study clearly rehabilitate dry minor mergers as feasible mechanisms for explaining partially their evolution.

\section{Conclusions} 
\label{Sec:conclusions}

Our collision-less N-body simulations show that intermediate and minor mergers are suitable mechanisms to drive bulge-growth along the S0 sequence early reported by \citet[][]{1976ApJ...206..883V} and recently updated by \citet[][]{2012ApJS..198....2K}. Different orbits, density ratios, and mass ratios (1:6, 1:9, 1:18) are considered, as well as two different models for the primary galaxy (S0b and S0c). 

We show that all remnants would be classified as undisturbed S0's by current specialized surveys. The merger events induce noticeable bulge growth in all cases, inducing the following morphological transformations: S0c$\longrightarrow$S0b and S0b$\longrightarrow$S0a. Depending on the initial $B/D$ of the primary galaxy that accretes the satellite, the merger induces bulge growth in two ways. In the case that the primary galaxy has a small bulge, the minor merger produces a prominent growth in $n$ and a noticeable rise in $B/T$ compared to their initial values, whereas in the case that it has a massive one, the increases in $n$ and $B/T$ are much more moderate for the same satellite. We also find that mergers with long pericentre orbits produce a much more moderate bulge growth than in analog cases with short impact parameters. In our experiments, the evolution in the $n$-$B/T$ plane depends more significantly on the orbit than on the mass ratio of the encounter for the same original S0 galaxy.

In all cases, the merger induces a negligible change of $r_\mathrm{e}$ and $h_\mathrm{D}$ in the galaxy, in such a way that the ratio of these two scale-lengths is nearly preserved. Therefore, if there is a coupling between the scale-lengths of the bulge and disk components prior to the merger, the merger does not break this coupling, but just disperses it slightly. This fact provides a simple explanation for the lack of correlation between the ratio of the bulge and disk scale-lengths and the S0 Hubble type reported by observations for S0's.  

Our models prove that dry intermediate and minor mergers onto S0 galaxies can make them to evolve within the sequence of S0 Hubble types towards earlier types, fulfilling global bulge-to-disk structural relations compatible with those observed in real S0's. This means that minor mergers should not be discarded from the evolution scenarios of S0's just on the basis of the strong correlations observed between the bulge and disk scale-lengths of these galaxies.

%
\small  
%
\begin{acknowledgements}   
The authors thank the anonymous referee for the provided input that helped to improve this publication significantly. We also thank Olga Sil'chenko for stimulating the present study, as well as Peter Erwin and Miguel Querejeta for interesting and useful comments. Supported by the Spanish Ministry of Science and Innovation (MICINN) under projects AYA2009-10368, AYA2006-12955, AYA2010-21887-C04-04, and AYA2009-11137, and by the Madrid Regional Government through the AstroMadrid Project (CAM S2009/ESP-1496, http://www.laeff.cab.inta-csic.es/projects/astromadrid/main/index.php). Funded by the Spanish MICINN under the Consolider-Ingenio 2010 Program grant CSD2006-00070: "First Science with the GTC" (http://www.iac.es/consolider-ingenio-gtc/). ACGG is a Ramon y Cajal Fellow of the Spanish MICINN. This research is based in part on services provided by the GAVO data center. It has made use of the NASA/IPAC Extragalactic Database (NED) which is operated by the Jet Propulsion Laboratory, California Institute of Technology, under contract with the National Aeronautics and Space Administration. S.~D.~H.~\& G.
\end{acknowledgements}

\bibliographystyle{aa}{}
\bibliography{elic0709_def}{}
\end{document}